# Domain wall dynamics in easy-cone magnets

Peong-Hwa Jang[1], Se-Hyeok Oh[2], and Kyung-Jin Lee[1,2,3†]

[1]Department of Materials Science and Engineering, Korea University, Seoul 02841, Korea

[2]Department of Nano-Semiconductor and Engineering, Korea University, Seoul 02841, Korea

[3]KU-KIST Graduate School of Converging Science and Technology, Korea University, Seoul 02841, Korea

[†]Correspondence to: K.-J.L. (kj_lee@korea.ac.kr)

**Abstract**

We theoretically and numerically investigate magnetic domain wall dynamics in a nanowire of easy-cone magnet. The easy-cone domain wall exhibits several distinguishing dynamic features in comparison to the easy-axis domain wall. The features of easy-cone domain wall are related to the generation of additional chiral spin textures due to the domain wall precession, which is common for various driving sources such as magnetic fields and spin-transfer torques. The unique easy-cone domain wall dynamics could enrich magnetic domain wall study and find use in device applications based on easy-cone domain walls.

# I. Introduction

A magnetic domain wall (DW) is a transient region between two domains, where the spin configuration continuously varies. Magnetic DWs can be used as information units in data storage and logic devices [1,2], demanding detailed understanding of DW dynamics. In this respect, the magnetic DW dynamics has been extensively studied for various classes of magnetic material such as ferromagnets [3-34], antiferromagnets [35,36], and ferrimagnets [37-43], and also for various driving means such as magnetic fields [3-8,37,38], spin-transfer torques [9-22], spin-orbit torques [23-27,35,36,39-43], and spin waves [28-34].

To date, most ferromagnetic DW studies have focused on ferromagnetic materials with the easy-axis state. In this work, we theoretically and numerically study DW dynamics in ferromagnetic materials with the easy-cone state. When the second-order magnetic anisotropy is non-negligible compared to the first-order one and satisfies a specific condition (described below), the equilibrium magnetization direction has an angle from the film normal. For the system where the cylindrical symmetry is preserved, the direction of equilibrium magnetization forms a cone with a finite angle, called the easy-cone state. Recently, the easy-cone magnet has attracted considerable interest for magnetic memories because of the short switching time and low switching current density [44-47] and for high-frequency oscillators because of its ability for zero-field oscillation [48]. Moreover, the easy-cone magnet is able to host spin superfluids associated with spontaneous breaking of the U(1) spin-rotational symmetry [49-54]. However, DW dynamics in the easy-cone magnet has not been investigated yet.

In this paper, we investigate DW dynamics induced by a magnetic field or a current in a conically magnetized nanowire. In section II, we introduce the easy-cone state and derive its equilibrium DW profile. In sections III and IV, we describe easy-cone DW dynamics induced by magnetic fields and spin-transfer torques, respectively. Lastly in section V, we show DW dynamics induced by spin injection at one side of the nanowire.

## II. Equilibrium easy-cone domain wall

Total magnetic energy of the system, including the exchange, first-order and second-order magnetic anisotropies, is given as

$$E_{\text{tot}} = \int dx \, [A_{\text{ex}}(\nabla\theta)^2 + K_{1,\text{eff}} \sin^2\theta + K_2 \sin^4\theta], \tag{1}$$

where $A_{\text{ex}}$ is the exchange stiffness constant, $K_{1,\text{eff}} = K_1 - 2\pi M_s^2$ is the first-order effective anisotropy energy density, $K_2$ is the second-order anisotropy energy density, $M_s$ is the saturation magnetization, and $\theta$ is the polar angle between the magnetization and $\hat{\mathbf{z}}$-axis (film normal). Figure 1(a) shows the phase diagram of magnetic state as a function of the first- and second-order magnetic anisotropies [55-57]. Perpendicular magnetization (perpendicular magnetic anisotropy: PMA) is stabilized when $K_{1,\text{eff}} > 0$ and $K_2 > -K_{1,eff}/2$. In-plane magnetization (easy plane) is stabilized when $K_{1,\text{eff}} < 0$ and $K_2 < -K_{1,\text{eff}}/2$. The PMA and easy-plane states coexist when $K_{1,\text{eff}} > 0$ and $K_2 < -K_{1,\text{eff}}/2$. Finally, the easy-cone state, which is of interest in this work, is stabilized when $K_{1,\text{eff}} < 0$ and $K_2 > -K_{1,\text{eff}}/2$.

For a single-domain easy-cone state, the energy minimization of Eq. (1) with respect to $\theta$ gives two equilibrium polar angles $\theta_{c1} = \sin^{-1}\sqrt{\kappa}$ and $\theta_{c2} = \pi - \sin^{-1}\sqrt{\kappa}$ where $\kappa = -K_{1,\text{eff}}/2K_2$. Figure 1(b) shows a schematic illustration of an easy-cone DW. With the magnetization $\mathbf{m} = (\cos\phi \sin\theta, \sin\phi \sin\theta, \cos\theta)$, the equilibrium one-dimensional DW profile of easy-cone magnet is derived by solving the Euler-Lagrange equation of the total magnetic energy [Eq. (1)] with the boundary conditions $\theta(x \to \infty) = \theta_{c1}$ and $\theta(x \to -\infty) = \theta_{c2}$, which is given as

$$\theta(x) = \tan^{-1}\left(\frac{\sqrt{1-\kappa}\tanh(\sqrt{\kappa(1-\kappa)}(x-X)/\lambda)}{\sqrt{\kappa}}\right) + \frac{\pi}{2}, \quad (2)$$

where $\lambda = \sqrt{A_{ex}/K_2}$ is the DW width and $X$ is the center position of the wall. In Fig. 1(c), we compare Eq. (2) with the DW profile obtained from numerical calculation with the following parameters: $A_{ex} = 1.2 \times 10^{-6}$ erg/cm, $K_{1,eff} = -3 \times 10^6$ erg/cm$^3$, $K_2 = 5 \times 10^6$ erg/cm$^3$, and $M_s = 1000$ emu/cm$^3$. We find a good agreement between Eq. (2) and modeling result. For a comparison, we also plot the DW profile of a PMA magnet ($A_{ex} = 1.2 \times 10^{-6}$ erg/cm, $K_1 = 7 \times 10^6$ erg/cm$^3$, $K_2 = 0$ erg/cm$^3$, and $M_s = 1050$ emu/cm$^3$) in Fig. 1(c).

Before ending this section, we note that there is an important difference in the equilibrium DW profile in between a PMA magnet and an easy-cone magnet. For a PMA DW, the in-plane magnetization component is zero at $x \to \pm\infty$ regardless of the azimuthal angle $\phi$ of the magnetization. In contrast, for an easy-cone DW, the in-plane component in the domain region varies depending on $\phi$ because $\sin\theta \neq 0$. This coupling between $\phi$ and the magnetization profile in the domain region results in unique dynamics of easy-cone DW when the DW precesses, as we will explain in the next section.

### III. Domain wall dynamics induced by magnetic field

Dynamics of easy-cone DW driven by a magnetic field applied in the $\hat{z}$ direction is studied by solving the Landau-Lifshitz-Gilbert (LLG) equation, given as

$$\frac{d\bm{m}}{dt} = -\gamma\, \bm{m} \times \bm{H}_{\text{eff}} + \alpha\, \bm{m} \times \frac{d\bm{m}}{dt}, \tag{3}$$

where $\gamma$ is the gyromagnetic ratio, $\bm{H}_{\text{eff}}$ is the effective magnetic field including the exchange, anisotropy, magneto-static, and external fields, and $\alpha$ is the Gilbert damping constant. Following Thiele's collective coordinate approach [58] for the DW position $X$ and DW angle $\varphi$ and with the equilibrium DW profile [Eq. (2)], we obtain the force equation as,

$$\alpha\left[\sqrt{(1-\kappa)\kappa} + (1-2\kappa)\tan^{-1}\left(\sqrt{\tfrac{1-\kappa}{\kappa}}\right)\right] \dot{X} + 2\sqrt{1-\kappa}\,\lambda(-\gamma\, H_z + \dot{\varphi}) = 0, \tag{4}$$

where $H_z$ is the magnitude of external field.

The steady state solution of the DW velocity ($\dot{\varphi} = 0$) is then given as,

$$v_{DW}(H_z) = \frac{2\sqrt{1-\kappa}}{\sqrt{(1-\kappa)\kappa} + (1-2\kappa)\tan^{-1}\left(\sqrt{\tfrac{1-\kappa}{\kappa}}\right)} \frac{\gamma \lambda H_z}{\alpha}. \tag{5}$$

Analytic solutions of the DW velocity beyond the steady state solution are difficult to obtain because the DW profile varies both spatially and temporally, as will be discussed below.

In the bottom panel of Fig. 2(a), we compare the analytic solution [Eq. (5), blue dotted line] of DW velocity with the velocity numerically calculated by micromagnetic simulations with $\alpha = 0.1$, nanowire length $L = 2$ μm, width $w = 50$ nm, and thickness $d = 0.8$ nm. Both one-dimensional (1D, blue diamond symbols) and 2D (black cross symbols, 25 discretized cells in the transverse direction of wire) simulations show similar results so that we focus on the 1D simulation results hereafter. In Fig. 2(a), we also show the DW velocity of a PMA magnet (black open symbols) for comparison. Numerical results for the easy-cone magnet are in agreement with Eq. (5) in low field regimes, whereas they largely deviate from Eq. (5) in high field regimes. This deviation is caused by the Walker breakdown [59], i.e., the DW precession above a threshold field. However, field dependence of DW velocity after the Walker breakdown is clearly different between the easy-cone magnet and the PMA magnet. Unlike the PMA DW that shows two separate regimes: steady motion below and precessional motion above a threshold field [see Fig. 2(a), black open symbols], the easy-cone DW shows three separate regimes: steady-state regime, intermediate regime, and precessional regime [see Fig. 2(a), solid diamond symbols]. In the intermediate regime, which is absent for the PMA DW, the velocity of the easy-cone DW shows several up and down jumps.

In the case of PMA DW, the precessional behavior after the Walker breakdown does not generate any additional magnetic textures in the uniform domain region and the azimuthal angle $\phi$ is spatially homogeneous even with the DW precession. In the case of easy-cone DW, however, the DW precession generates additional magnetic textures because the easy-cone state stabilizes nonzero in-plane component of magnetization in the domain region. In Fig. 2(b), we schematically describe two kinds of easy-cone DW profiles. In the absence of magnetic field [the upper panel of Fig. 2(b)], the polar angle $\theta$ of magnetizations follows Eq. (2) and the azimuthal angle $\phi$ is $\pi$ to minimize the exchange and shape anisotropy energies. When applying a magnetic field below a threshold for the precession, $\phi$ at the DW center tilts from $\pi$ a little bit, which in turn changes $\phi$ of magnetizations near the DW to reduce the exchange energy. Even in this case, $\phi$ at $x \to \pm\infty$ is still $\pi$ because of the shape anisotropy. As a result, $\phi$ is no longer constant and becomes inhomogeneous. When applying a magnetic field above a threshold, $\phi$ at the DW center rotates by about $\pi$ and is then close to 0 [the bottom panel of Fig. 2(b)]. In this situation, $\phi$ changes from $\pi$ at $x \to -\infty$, through $\approx 0$ at $x = 0$, to $\pi$ at $x \to +\infty$. Because of the $\pi$-rotation of $\phi$ at the DW center, two inhomogeneous magnetic textures are formed at the front and rear sides of DW. We call this magnetic texture originating from the $\pi$-rotation of $\phi$ as a sub-DW (SD).

In Fig. 2(c), we show a top view of simulated magnetization configuration in the easy-cone nanowire for $H_z = 60$ Oe above the threshold for the Walker breakdown. In this case, a SD is formed at the front of DW and another SD is formed at the rear of DW. With increasing a magnetic field, the DW center magnetization undergoes more $\pi$-rotations of $\phi$ and as a result, more SDs are generated. On the top panel in Fig. 2(a), we show the number of SDs created at the front of DW as a function of $H_z$. For magnetic fields in the intermediate regime (50 Oe $<H_z<$ 150 Oe), the number of SDs increases discontinuously. This discontinuity is caused by the fact that for an additional $\pi$-rotation of $\phi$ of the DW center magnetization, a corresponding exchange (and other) energy cost must be overcome and that the new state is stabilized by the shape anisotropy of nanowire. Whenever an additional SD is created, the DW velocity shows a discontinuous drop [indicated by a black arrow in Fig. 2(a)]. We attribute this velocity drop to the fact that the magnetic field must move not only the DW but also the additional SD acting as an additional energy barrier. When increasing the magnetic field, the DW velocity increases again until the field is high enough to create the next SD.

On the other hand, the number of SDs at the front of DW decreases at higher fields corresponding to the precessional regime [$H_z>$ 150 Oe; the top panel of Fig. 2(a)]. Whenever the magnetization at the DW center undergoes a $\pi$-rotation, it is obvious that the number of SDs generated at the rear of DW increases. We observe that the number of SDs at the front side, however, does not increase continuously with the field because the distance among the front SDs decreases as the source of SD generation (i.e., precessing DW) moves faster towards the front side. The decreased distance among SDs makes the magnetic texture energetically unfavorable, which limits the creation of SDs. The newly created SD exceeding the limited maximum number, collapses and merges with the last one, resulting in the decreased SD number at the front side. We also remark that because of its unique feature of field-driven easy-cone DW dynamics in the presence of the shape anisotropy, multiple SDs remain in the nanowire even after the field is switched off.

### IV.   Domain wall dynamics induced by spin-transfer torque

We investigate easy-cone DW dynamics induced by adiabatic and nonadiabatic spin-transfer torques (STTs) [13-15], which is described with the modified LLG equation as

$$\frac{d\bm{m}}{dt} = -\gamma\,\bm{m}\times\bm{H}_{eff} + \alpha\,\bm{m}\times\frac{d\bm{m}}{dt} + b_{\mathrm{j}}\bm{m}\times\left(\bm{m}\times\frac{d\bm{m}}{dx}\right) + \beta b_{\mathrm{j}}\left(\bm{m}\times\frac{d\bm{m}}{dx}\right). \quad (6)$$

Here $b_{\mathrm{j}} = \hbar P J_{\mathrm{e}}/2eM_s$ is the spin current velocity, $\hbar$ is the reduced Plank constant, $P$ is the spin polarization, $J_{\mathrm{e}}$ is the current density, $e$ is the electric charge, and $\beta$ is the non-adiabaticity. Following Thiele's approach, the steady-state solution of easy-cone DW driven by STT gives $v_{\mathrm{DW}} = \gamma\beta b_{\mathrm{j}}/\alpha$, identical to that of PMA DW [14,15]. We show the average velocity over a time period as a function of $\beta$, in comparison to numerical results in Fig. 3(a).

When $\beta \neq \alpha$, the easy-cone DW shows processional motion above a threshold current density. We note that once the easy-cone DW precesses, the SD generation is a common feature regardless of type of the driving source. Therefore, the STT is also able to create SDs. Figure 3(b) shows temporal evolution of DW velocity (bottom panel), the number of front SDs (middle panel), and magnetization configuration (top panel; top view), induced by STT above a threshold. In contrast to the field-driven generation of front SDs of which maximum number is limited, the front SDs are continuously generated when driven by STT even right above a threshold ($J_e = 4.8 \times 10^7$ A/cm$^2$). The continuous increase in the number of front SDs in STT-driven case is caused by the fact that STT induces dynamics of SDs as well as DW in the same direction because both types of spin texture have finite spatial gradients and thus experience non-zero STT. Hence the DW induced by STT does not feel the SD as an additional barrier that hinders its dynamics. Once a SD is created, similarly, there is a sudden drop of velocity [indicated by a blue arrow in bottom panel of Fig. 3(b)] but recovers its velocity shortly.

V.   **Domain wall dynamics induced by spin injection at wire edge.**

In this section, we show DW dynamics driven by spin injection at one side of a nanowire [see Fig. 4(a)]. The coexistence of spin superfluidity and a DW in easy-cone state originates from the fact that the ground states of easy-cone magnet break U(1) and Z2 symmetries simultaneously. In Ref. [53], Kim *et al.* theoretically investigated easy-cone DW dynamics by spin injection with neglecting the non-local magnetostatic coupling, resulting in the shape anisotropy along the wire-length direction. They found that the easy-cone DW moves along a particular direction regardless of the magnitude of injected spin current. In this section, we investigate how the shape anisotropy, which is usually uneasy to remove from realistic nanowires, changes this DW dynamics. We show below that the easy-cone DW can move in an opposite direction, hence, change its velocity sign when we take into account the shape anisotropy via an additional anisotropy in the $\hat{x}$-direction.

To get an insight into DW dynamics by spin injection, two terms are additionally considered to Eq. (1), the exchange energy with respect to $\phi$ and the simplified non-local magneto-static energy in a nanowire, which gives a modified total magnetic energy $E_{tot}'$ as,

$$E_{tot}' = \int (A_{ex}\{(\nabla\theta)^2 + \sin^2\theta\,(\nabla\phi)^2\} + K_{1,eff}\sin^2\theta + K_2\sin^4\theta + K_D\cos^2\theta\sin^2\phi)dx. \quad (7)$$

Here we include the exchange, first- and second-order anisotropy energy, and shape anisotropy energy ($K_D$). DW dynamics can be interpreted by employing equations of motion of the system in the spherical coordinate, given as

$$\alpha s \dot{\theta} - s\sin\theta\,\dot{\phi} = \frac{\partial E_{tot}'}{\partial \theta}, \quad (8a)$$

$$s\sin\theta\,\dot{\theta} + \alpha s\sin^2\theta\,\dot{\phi} = \frac{\partial E_{tot}'}{\partial \phi}, \quad (8b)$$

where $s = -\frac{M_s}{\gamma}$ is the spin moment. For the moving frame with wall velocity $v_{SI}$, we reorganize Eq. (8) to the linear order in $v_{SI}$,

$$-s\sin\theta\,\omega + \alpha s(-v_{SI}\nabla\theta)$$
$$= A_{ex}\left(-\nabla^2\theta + \frac{K_2}{A_{ex}}\partial_\theta(\sin^2\theta - \sin^2\theta_c) - \frac{2K_D}{A_{ex}}\sin\theta\cos\theta\sin^2\phi\right), \quad (9a)$$

$$s\sin\theta\,(-v_{SI}\nabla\theta) + \alpha s\sin^2\theta\,\omega = A_{ex}\left(-\nabla(\sin^2\theta\,\nabla\phi) + \frac{2K_D}{A_{ex}}\cos^2\theta\sin\phi\cos\phi\right), \quad (9b)$$

where $\omega \equiv \dot{\phi}$, $\theta_c \equiv \sin^{-1}\sqrt{-\frac{K_{1,eff}}{2K_2}}$, $\nabla\theta \equiv \partial_x\theta$, and $\nabla\phi \equiv \partial_x\phi$.

We note that spin injection into the conically magnetized nanowire generates SDs at the source and hence $\phi$ is not a constant both temporally and spatially. This spatiotemporal variation of $\phi$ makes the integration of Eq. (9) impossible. In order to make Eq. (9) integrable, we use a crude assumption that $\phi$ is spatially uniform and obtain the following equations for $v_{SI}$ and $\omega$:

$$2s \cos\theta_c \, \omega + \alpha\, s\, A_v v_{SI} = 0, \tag{10a}$$

$$2\, s\, \cos\theta_c \, v_{SI} - \alpha\, s\, A_{\omega 1}\, \omega = j_s - 2K_D\, A_{\omega 2} \sin\phi \cos\phi, \tag{10b}$$

where $A_v = \frac{\sin 2\theta_c + (\pi - 2\theta_c)\cos 2\theta_c}{2\lambda}$, $A_{\omega 1} = l \sin^2\theta_c + \lambda(\pi - 2\theta_c)$, $A_{\omega 2} = l \cos^2\theta_c - \lambda(\pi - 2\theta_c)$, and $j_s = -A_{ex} \sin^2\theta_c \nabla\phi = (\hbar P\, j_{inj}\, d_{inj})/(2\, e\, t_{EC})$ is the spin current generated from the source. Here, $l$ is the distance between the DW center and the injection source, $j_{inj}$ is the current density injected from a ferromagnet (FM), $d_{inj}$ is the length of FM on $\hat{x}$-direction, and $t_{EC}$ is the thickness of easy-cone nanowire [see Fig. 4(a)]. From Eq. (10), we derive the DW velocity $v_{SI}$ by spin injection, given as

$$v_{SI} = \frac{\gamma\left(2K_D\, A_{\omega 2} \sin\phi \cos\phi - \frac{\hbar P\, j_{inj}\, d_{inj}}{2\, e\, t_{EC}}\right)}{M_s \left(2\, \cos\theta_c + \frac{\alpha^2 A_{\omega 1} A_v}{2 \cos\theta_c}\right)}. \tag{11}$$

One finds from Eq. (11) that $v_{SI}$ is either positive or negative depending on $\phi$ and $j_{inj}$. The threshold current density $j_{th}$ below and above which the sign of $v_{SI}$ changes can be obtained by maximizing $\sin\phi \cos\phi$ in the numerator of Eq. (11) and setting $v_{SI}=0$, given as

$$j_{th} = \frac{2eK_D A_{\omega 2} t_{EC}}{\hbar P\, d_{inj}}. \tag{12}$$

In Fig. 4(a), we show a schematic view of the system containing an easy-cone DW at the center ($i = L/2$) of the nanowire with the spin injection source (FM1) located at $i = L/4$ (source area of $20 \times 50$ nm$^2$). We perform micromagnetic simulations with the following parameters: $K_{1,eff} = -3 \times 10^6$ erg/cm$^3$, $K_2 = 5 \times 10^6$ erg/cm$^3$, $M_s = 1000$ emu/cm$^3$, polarization factor $P = 0.3$, $\alpha = 0.1$, and injection length $d_{\text{inj}} = 20$ nm. Spins are injected for a duration of 100 ns. Spin injection induces magnetization precession at the injection area and generates spin current proportional to $A_{\text{ex}} \sin^2 \theta \, \nabla \phi$, which propagate and eventually interact with the DW. In Fig. 4(b), we plot the time evolution of DW velocity, which shows two different DW motion with opposite sign, depending on the injected current density. Corresponding top views of magnetization configuration for low and high current densities are shown in Fig. 4(c) and (d), respectively.

For an injected current below a threshold [$j_{\text{th}}$, Eq. (12)], the spin current is unable to precess the easy-cone DW due to the shape anisotropy and thus cannot generate SDs at the front side of DW. In this case, the spin current (or, equivalently, SDs at the rear side of DW) just pushes the DW, resulting the DW motion along the direction of spin current flow (i.e., a positive $v_{\text{SI}}$ in our sign convention) [see Fig. 4(c)]. On the other hand, for an injected current above $j_{\text{th}}$, the spin current induces DW precession and its angular momentum is transferred to the DW as explained in Ref. [53]. In this case, the wall velocity shows an oscillatory behavior with a negative sign in average and pulls the DW towards the spin current source [see Fig. 4(d)]. For a comparison, we estimate a theoretical $j_{\text{th}}$ from Eq. (12) using the same parameters for simulations and obtain $j_{\text{th}} \approx 7 \times 10^7$ A/cm$^2$, which is smaller than the numerically obtained value ($\approx 11 \times 10^7$ A/cm$^2$), possibly due to the crude approximation adopted to derive Eq. (12). Despite somewhat unsatisfactory quantitative agreement, we note that Eqs. (11) and (12) reveal the underlying mechanism of the sign change in $v_{\text{SI}}$, depending on the magnitude of injected current. Usually, to change the DW motion direction during the device operation, one has to use a transistor that supplies bipolar currents. The above-mentioned bidirectional easy-cone

DW motion induced by a *unipolar* current would be useful for device applications as one can use a less expensive diode rather than a transistor.

## VI. Summary

We have investigated easy-cone DW dynamics induced by a magnetic field or an electric current in conically magnetized nanowires. We find the easy-cone DW dynamics is closely related to the generation of sub-DWs caused by the combined action between the easy-cone ground state and DW precession. For a field-driven case, the sub-DW generation results in unique intermediate regime where the DW velocity shows several up and down jumps. For a STT-driven case, this intermediate regime is absent because STT moves not only the DW but also the sub-DWs in the same direction. Lastly, for a spin injection case, we find that the DW motion direction can be controlled by varying the injected current density. Our work will provide a guideline for an experimental study on the easy-cone DW dynamics with various driving sources.

**Acknowledgement**

We acknowledge Se Kwon Kim for fruitful discussion. This work was supported by the National Research Foundation of Korea (NRF) (Grants No. 2015M3D1A1070465 and No. 2017R1A2B2006119) and the KIST Institutional Program (Project No. 2V05750).

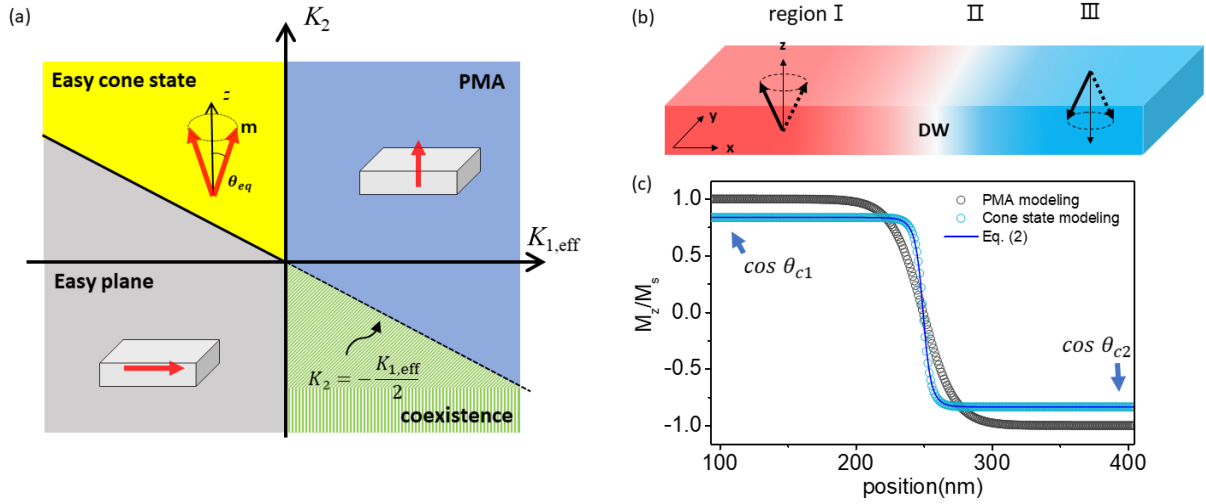

Figure. 1. Easy-cone state and its wall profile. (a) Phase diagram of magnetic state as a function of effective first-order anisotropy $K_{1\text{eff}}$ and second-order anisotropy $K_2$. (b) Schematic illustration of conically magnetized nanowire. Regions I and III are uniform domain parts where the equilibrium polar angles of magnetization are $\theta = \sin^{-1}\sqrt{-\frac{K_{1,\text{eff}}}{2K_2}}$ and $\pi - \sin^{-1}\sqrt{-\frac{K_{1,\text{eff}}}{2K_2}}$, respectively, whereas region II is domain wall part. (c) Wall profile represented as normalized $M_z$ component of easy-cone (blue open symbols) and PMA (black open symbols) domain wall.

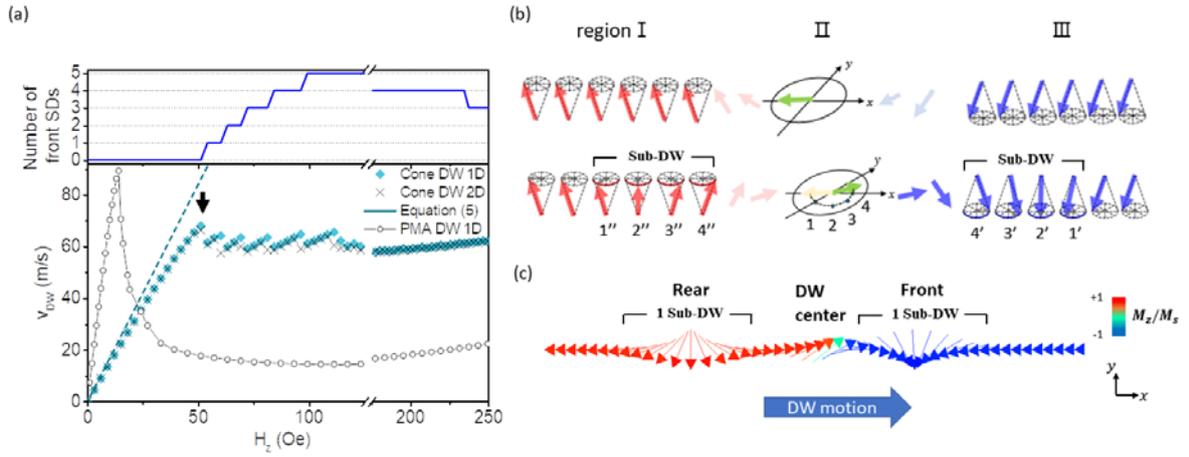

Figure 2. Field-induced domain wall motion in a conically magnetized nanowire. (a) Domain wall velocity (bottom panel) and the number of SD (top panel) as a function of applied magnetic field. (b) Schematic illustration of easy-cone domain wall profile in the absence (top) and presence (bottom) of magnetic field. (c) Top view of magnetization configuration for $|H_z| = 60$ Oe. Color code represents the z-component of magnetization, $M_z$.

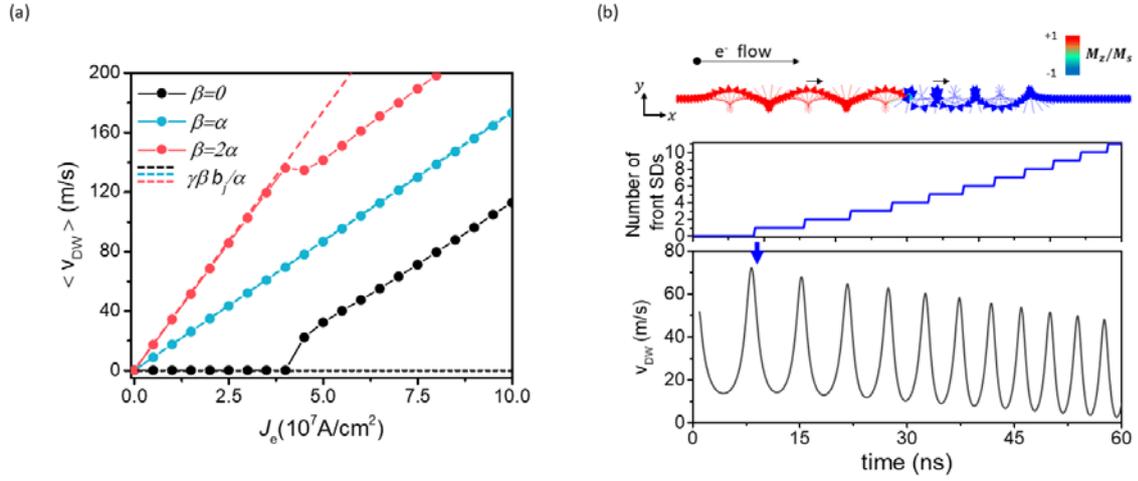

Figure 3. Spin-transfer torque induced domain wall motion in a conically magnetized nanowire. (a) Domain wall velocity as a function of applied current for various non-adiabaticities $\beta = 0, \alpha,$ and $2\alpha$. (b) Top view of magnetization configuration (top panel) and time evolution of domain wall velocity and the number of SD (bottom panel) for $J_e$=4.8×10$^7$ A/cm$^2$. Color code represents the z-component of magnetization, $M_z$. We use the same parameters of Fig. 1 and $P = 0.3$.

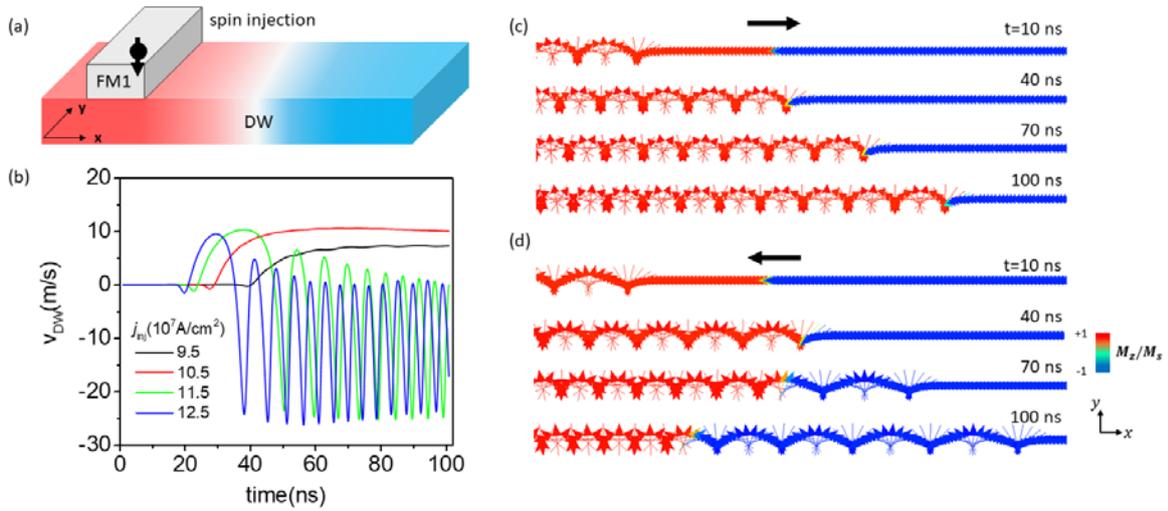

Figure 4. Domain wall motion in a conically magnetized nanowire, induced by local spin injection. (a) Schematic illustration of local spin injection in a conically magnetized nanowire. (b) Domain wall velocity as a function of time for various current densities. Easy-cone DW moving (c) away from the injection source (positive velocity) for $j_{inj} = 10.5 \times 10^7 $A/cm$^2$ and (d) towards the source (negative velocity in average) for $j_{inj} = 11.5 \times 10^7$A/cm$^2$. Color code represents the $M_z$ component of magnetization.